\documentclass[pdflatex,sn-mathphys-num]{sn-jnl}

\usepackage{graphicx}%
\usepackage{multirow}%
\usepackage{amsmath,amssymb,amsfonts}%
\usepackage{amsthm}%
\usepackage{mathrsfs}%
\usepackage[title]{appendix}%
\usepackage{xcolor}%
\usepackage{textcomp}%
\usepackage{manyfoot}%
\usepackage{booktabs}%
\usepackage{listings}%
\usepackage{algorithm2e}
\usepackage{algpseudocode}
\usepackage{array}
\usepackage{placeins}
\usepackage{subfig}

\usepackage{makecell}

\theoremstyle{thmstyleone}%
\theoremstyle{thmstyletwo}%

\theoremstyle{thmstylethree}%

\begin{document}

\title[Article Title]{Promptable segmentation with region exploration enables minimal-effort expert-level prostate cancer delineation}

\author[1]{\fnm{Junqing} \sur{Yang}}

\author[2]{\fnm{Natasha} \sur{Thorley}}

\author[3]{\fnm{Ahmed Nadeem} \sur{Abbasi}}

\author[2]{\fnm{Shonit} \sur{Punwani}}

\author[4]{\fnm{Zion} \sur{Tse}}

\author[1]{\fnm{Yipeng} \sur{Hu}}

\author*[1,4]{\fnm{Shaheer U.} \sur{Saeed}}\email{shaheer.saeed@qmul.ac.uk}

\affil[1]{\orgdiv{UCL Hawkes Institute; Department of Medical Physics and Biomedical Engineering}, \orgname{University College London}, \orgaddress{UK}}

\affil[2]{\orgdiv{Centre of Medical Imaging}, \orgname{University College London}, \orgaddress{UK}}

\affil[3]{\orgdiv{Department of Oncology}, \orgname{Aga Khan University Hospital}, \orgaddress{Pakistan}}

\affil[4]{\orgdiv{Centre for Bioengineering; School of Engineering and Materials Science}, \orgname{Queen Mary University of London}, \orgaddress{UK}}


\abstract{\textbf{Purpose:} 
Accurate segmentation of prostate cancer on magnetic resonance (MR) images is crucial for planning image-guided interventions such as targeted biopsies, cryoablation, and radiotherapy. However, subtle and variable tumour appearances, differences in imaging protocols, and limited expert availability make consistent interpretation difficult. While automated methods aim to address this, they rely on large expertly-annotated datasets that are often inconsistent, whereas manual delineation remains labour-intensive. This work aims to bridge the gap between automated and manual segmentation through a framework driven by user-provided point prompts, enabling accurate segmentation with minimal annotation effort.

\textbf{Methods:} 
The framework combines reinforcement learning (RL) with a region-growing segmentation process guided by user prompts. Starting from an initial point prompt, region-growing generates a preliminary segmentation, which is iteratively refined through RL. At each step, the RL agent observes the image and current segmentation to predict a new point, from which region growing updates the mask. A reward, balancing segmentation accuracy and voxel-wise uncertainty, encourages exploration of ambiguous regions, allowing the agent to escape local optima and perform sample-specific optimisation. Despite requiring fully supervised training, the framework bridges manual and fully automated segmentation at inference by substantially reducing user effort while outperforming current fully automated methods.

\textbf{Results:} 
The framework was evaluated on two public prostate MR datasets (PROMIS and PICAI, with 566 and 1090 cases). It outperformed the previous best automated methods by 9.9\% and 8.9\%, respectively, with performance comparable to manual radiologist segmentation, reducing annotation time tenfold.

\textbf{Conclusion:} 
By combining prompting with RL–driven exploration, the framework achieves radiologist-level prostate cancer segmentation with a fraction of the annotation effort, highlighting the potential of RL to enable adaptive and efficient cancer delineation.\\
Code: \url{github.com/JQ-Sakura/prostate-rl-segmentation}
}



\maketitle

\section{Introduction}\label{sec:intro}

Magnetic resonance (MR) imaging plays a central role in interventional planning, for both diagnosis and treatment of prostate cancer \cite{Ahmed2017PROMIS}. However, accurate interpretation of prostate MR images is particularly challenging due to the variable and often subtle imaging appearances of cancerous tissue. This is echoed by the low reported sensitivity, even in expert readings \cite{Ahmed2017PROMIS}. The task is further complicated by differences in acquisition protocols, imaging equipment, and radiologist training \cite{Penzkofer2014NMRB}. Reliable assessment therefore requires specialist expertise, which is severely limited, especially in resource-constrained regions \cite{DosSantosSilva2022CommunMed}. Even when expertise is available, substantial annotation burden and inter-observer variability persist, reflecting the inherent difficulty of the task \cite{DosSantosSilva2022CommunMed}. This, combined with the significant manual effort required for such assessment, leads to inconsistent reporting, which undermines the potential of MR imaging to guide timely and effective interventional procedures \cite{Ahmed2017PROMIS}.

Automation has been widely explored to reduce diagnostic variability and accelerate reporting workflows, particularly through automated tumour boundary delineation or segmentation \cite{Sanders2022RadOnc_PartII}. However, fully automated systems typically depend on large, meticulously labelled datasets for supervised training \cite{Sunoqrot2022ERRE}. Curating such datasets is hampered by the limited availability of experts, the significant time burden of delineating cancerous tissue, and high inter-annotator variability \cite{Sunoqrot2022ERRE}. As a result, automated models often inherit biases and inconsistencies from their training data, leading to unreliable performance in clinical practice \cite{Sanders2022RadOnc_PartII}.

Semi-automated methods seek to mitigate some of the variability and burden of manual segmentation, by incorporating user guidance into the automated segmentation process \cite{Karam2025Promptable,Ma2024MedSAM}. Recently, promptable methods have shown promise for various vision tasks, where user inputs, such as point or bounding box prompts, guide segmentation \cite{Kirillov2023SAM}. This is distinct from interactive segmentation where users may need to refine segmentation masks iteratively. In contrast, promptable methods necessitate only one, or a few, prompts to initiate automated segmentation. Most promptable models are developed for general vision or anatomical segmentation tasks \cite{Kirillov2023SAM,Ma2024MedSAM}, where image variability is relatively low compared to pathological cancer segmentation. These approaches are typically built on conventional deep learning frameworks, which primarily learn global dataset-level patterns making them prone to converging toward local optima when datasets are limited or imperfect \cite{Aggarwal2021NPJDM,Karam2025Promptable}. For prostate cancer, where appearance variability is high and gathering sufficiently-sized datasets is challenging, such dataset-level strategies rarely achieve clinically acceptable performance \cite{Aggarwal2021NPJDM,Yan2024Combiner}.

\begin{figure}[!ht]
    \centering
    \includegraphics[width=0.7\linewidth]{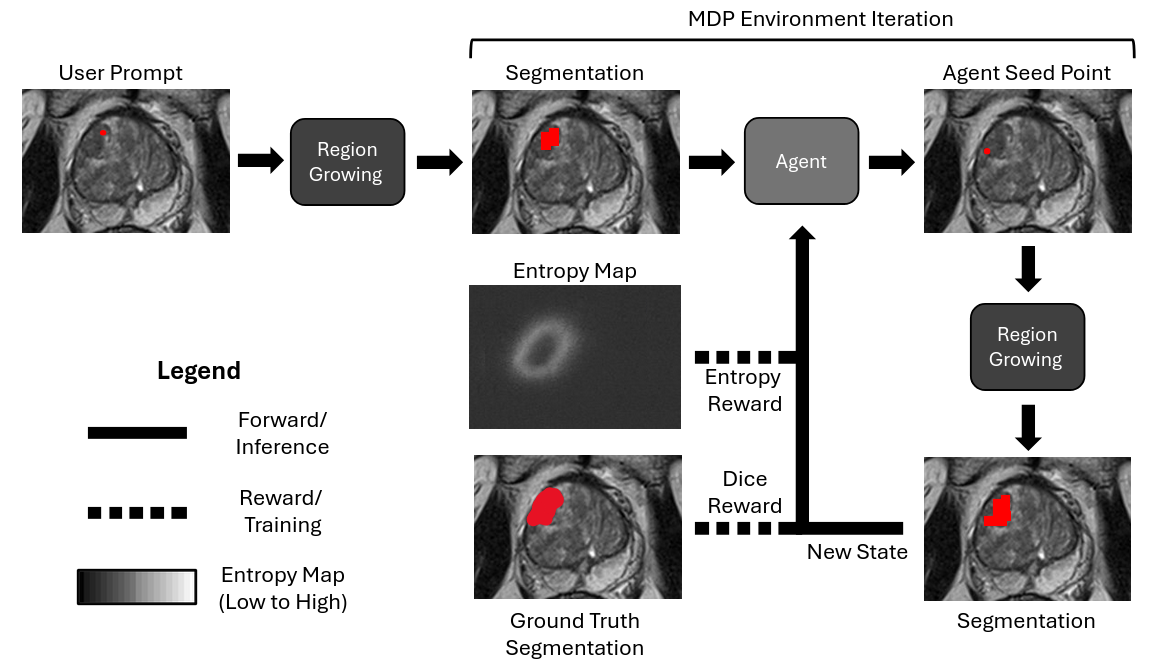}
    \caption{An overview of the proposed promptable segmentation using RL.}
    \label{fig:method}
\end{figure}

To overcome these limitations, we propose a reinforcement learning (RL)-based promptable segmentation framework tailored for prostate cancer on MR images. The method formulates segmentation as a dynamic sequential process, where a RL agent iteratively refines a segmentation mask guided by user-provided point prompts. Each prompt initiates a region-growing module \cite{Adams1994SRG} that produces an initial segmentation of the region of interest (ROI). The agent then observes both the image and current segmentation to predict a new seed location expected to improve the result, from which region growing is re-applied to update the mask. Reinforcement learning enables this process to incorporate exploration, guided by voxel-wise entropy, allowing the agent to search challenging or uncertain regions and escape dataset-level local optima arising from limited or imperfect data \cite{Aggarwal2021NPJDM, Czolbe2021IsUncertainty, Ahn2022PerSampleGrad, saeed2024competing}. Through this exploration–exploitation balance, the model can efficiently search for and converge on optimal sample-specific segmentation solutions, even in challenging high variability cases. The framework is summarised in Fig. \ref{fig:method}. At inference time, the proposed framework bridges the gap between manual and fully automated segmentation by approaching expert-level accuracy with only sparse user interaction, substantially reducing annotation effort compared to full manual delineation. Fully supervised annotations are still required during training; the primary benefit lies in reducing expert effort during deployment rather than dataset curation.

RL has previously been applied in medical imaging, for example for landmark localisation, where agents are trained to identify fixed anatomical points through sequential search \cite{vlontzos2019multiple, alansary2019evaluating}. Segmentation, however, is fundamentally different, as it requires refinement of spatially extended region-of-interest boundaries rather than localisation of a single target. In this work, RL is introduced to explicitly incorporate exploration–exploitation into promptable segmentation, enabling the agent to escape dataset-level local optima by actively exploring uncertain regions on a per-sample basis during inference. This capability is particularly important for pathological segmentation, where tumour appearances are heterogeneous and training data are limited, making single-pass foundation and supervised models, that rely on dataset-level trends, prone to failure on atypical cases.


The contributions of this work are summarised: 1) developing a novel promptable segmentation mechanism, using RL to allow sample-specific optimisation, for prostate cancer segmentation on MR images; 2) evaluating the proposed method for prostate cancer segmentation using two publicly available datasets of 566 and 1090 prostate cancer patient MR images; 3) demonstrating performance that substantially exceeds recent state-of-the-art methods such as nnUNet, UNeTr and Combiner; 4) demonstrating performance that approaches the level of expert radiologists while requiring minimal annotation effort at inference; 5) open-source implementation available at: \url{github.com/JQ-Sakura/prostate-rl-segmentation}.

\section{Methods}\label{sec:methods}


\subsection{Definitions for the image, voxel, and segmentation}

Let $x\in\mathcal{X} \subseteq \mathbb{R}^{H \times W \times D}$ denote a volumetric image with spatial dimensions $(H, W, D)$, where $\mathcal{X}$ is the space of images. Its corresponding segmentation mask is defined as $y\in\mathcal{Y} \subseteq \{0,1\}^{H \times W \times D}$, where $\mathcal{Y}$ is the space of segmentation masks. Each voxel is indexed by $v \in \mathcal{V} \subseteq \mathbb{Z}^{3}$, where $\mathcal{V}$ is the space of voxel indices and $x(v)$ denotes the intensity at voxel $v$. The goal of the framework is to generate an optimal segmentation that delineates the prostate cancer region within the image.

\subsection{Efficient exploration of segmentation using region growing}

We define a region-growing operator that expands a segmentation from a single seed voxel.
Let the seed be $v_s \in \mathcal{V}$. The region-growing mapping is written as
$g(\cdot):\ \mathcal{X}\times \mathcal{V} \rightarrow\ \mathcal{Y}$, where for a specific sample the segmentation from region-growing is given by $y= g(x, v_s)$. The region-growing is itself considered non-parametric in our formulation as any development and parameter tuning happens prior to the promptable segmentation outlined in Sec. \ref{sec:rl-promptseg}

\paragraph{Surrogate segmentation network and entropy}\label{subsec:surrogate-entropy}

A surrogate neural network $f(\cdot;\theta):\mathcal{X}\rightarrow[0,1]^{H \times W \times D}$ parameterised by weights $\theta$, is employed to provide voxel-wise probabilities that guide the region-growing process. For a given image $x \in \mathcal{X}$, the network produces a probability map $y_p = f(x;\theta)$, where $y_p\in[0,1]^{H \times W \times D}$. 

While $y_p$ encodes model confidence, it does not explicitly capture uncertainty.
To quantify voxel-wise uncertainty, an entropy map $y_e$ is derived from $y_p$ as $y_e(v)=-[y_p(v)\log y_p(v) + (1-y_p(v))\log(1-y_p(v))]$. Low-entropy voxels correspond to confident predictions (probabilities near 0 or 1), while high-entropy voxels indicate ambiguity (probabilities near 0.5).
The surrogate network $f$ always outputs a voxel-wise probability map $y_p$, from which an entropy map $y_e$ is deterministically derived. For brevity, this derivation is summarised simply as $y_e=f(x;\theta)$, which is a notational shorthand rather than a redefinition of the network output.

The entropy at each voxel is then given by $y_e(v)$, which measures the uncertainty for that particular voxel.

The surrogate network is trained separately in a fully supervised manner using a set of annotated training samples $\{x_i, y_i\}_{i=1}^{N}$, where $x_i \in \mathcal{X}$ and $y_i \in \mathcal{Y}$. The training objective minimises a loss function $\mathcal{L}:\mathcal{Y}\times\mathcal{Y}\rightarrow\mathbb{R}$, where the loss for a particular sample is given by $\mathcal{L}(f(x_i;\theta), ~y_i)$, which is the Dice loss in our work. The optimisation to obtain optimal parameters $\theta^*$ is then conducted using:

\begin{equation}
    \theta^* = \arg \min_{\theta} \frac{1}{N} \sum_{i=1}^N \mathcal{L}(f(x_i;\theta), ~y_i)
    \label{eq:dice-loss}
\end{equation}

After optimisation, parameters $\theta^*$ remain fixed and $f(\cdot;\theta^*)$ is used to generate entropy maps $y_e=f(\cdot;\theta^*)$ for the region-growing process.
The surrogate network is trained independently and remains fixed during RL training and inference. In future work, the use of pre-trained surrogate models could be explored to reduce computational overhead.

\subsubsection{Region expansion for segmentation}

In this subsection, $j$ denotes the region-growing iteration index.
The segmentation $y$ is initialised using the seed voxel $v_s$:
\[
y_{j=0}(v) =
\begin{cases}
1, & v = v_{s},\\
0, & \text{otherwise.}
\end{cases}
\]

Each voxel with $y_j(v)=1$ is considered included in the current ROI segmentation, where $v'$ denotes an included voxel.
For each included voxel $v'$, the neighbourhood is defined as the set of voxels in a local window around $v'$, expressed as $v = v' + \Delta v$ where $\Delta v = (\Delta a, \Delta b, \Delta c)$ controls neighbourhood radius (set as $(3,3,3)$ in our work).

For each candidate voxel $v$ in the neighbourhood $v' + \Delta v$ that is not yet included (i.e., $y_j(v)=0$), the inclusion condition is:
\[
y_{j+1}(v) =
\begin{cases}
1, & \text{if } \sigma_x(v) < \tau_\sigma \text{ and } y_e(v) < \tau_e,\\[3pt]
y_j(v), & \text{otherwise.}
\end{cases}
\]
where $\sigma_x(v)$ is the neighbourhood intensity standard deviation computed over all voxels within $v \pm \Delta v$, and $y_e(v)$ is the voxel-wise entropy obtained from $f(x; \theta^*)$. Both $\tau_\sigma$ and $\tau_e$ are hyper-parameters which are left unchanged from defaults in \cite{Adams1994SRG}.

This process repeats iteratively, where each iteration examines the neighbours of newly included voxels.
Region growing terminates when no new voxels are added, i.e. $\|\, y_{j+1} - y_j \,\|_1 = 0$,
or when a predefined maximum number of iterations $J$ is reached.
Final segmentation after convergence is denoted as $y = y_J$, at final iteration index $J$.

The entire region-growing operator can be summarised as $y= g(x, v_s)$, with $y$ being the final segmentation mask, $x$ the image and $v_s$ being the initial seed point.

\subsection{Reinforcement learning for promptable segmentation}\label{sec:rl-promptseg}

RL is used to model promptable segmentation as a sequential process, where an agent iteratively refines the segmentation by selecting new seed points for region growing.

A neural network agent $h(\cdot; \phi): \mathcal{S} \rightarrow \mathcal{A}$, parameterised by weights $\phi$, defines the policy that maps the observed state $s_t\in\mathcal{S}$ at time step $t$ to the next action $a_t\in\mathcal{A}$, where $\mathcal{S}$ and $\mathcal{A}$ are the state and action spaces. \\

\noindent\textbf{State:} The state observed by the agent at time step $t$ is defined as $s_t\in\mathcal{S}$, which consists of $x$ the MR image and $y_t$ the current segmentation mask after the $t$-th step. This means that $s_t=(x,y_t)$ and that $\mathcal{S}=\mathcal{X}\times \mathcal{Y}$.\\

\noindent\textbf{Action:} At each time step $t$, the agent selects an action $a_t \in \mathcal{A}$, which corresponds to the next seed voxel location to initialise region growing.
Formally, the action is given by $a_t=v_{s,t}=h(s_t;\phi)$, which means that $\mathcal{A}=\mathcal{V}$. \\

\noindent\textbf{State transition:} After the agent selects an action $a_t = v_{s,t}$, the environment updates the segmentation by applying the region-growing operator. The new state is denoted as $s_{t+1} = (x, y_{t+1})$, where $y_{t+1} =  g(x, v_{s,t})$ is obtained from the region-growing operator. This completes one transition in the Markov decision process transitioning from the current state-action pair $(s_t, a_t)$ to the next state $s_{t+1}$.\\

\noindent\textbf{Reward:} The agent receives a scalar reward that quantifies the improvement in segmentation after each action. For a given time step $t$, the reward function is defined as $r:\mathcal{S}\times\mathcal{A}\rightarrow\mathbb{R}$, and the reward at time step $t$ is given by:

\begin{align}
R_t = r(s_t, a_t) = \mathcal{L}(y_t, \hat{y})  - \mathcal{L}(y_{t+1}, \hat{y}) 
+ \beta\, \mathbb{E}_{v \in y_{t+1}}[y_e(v)],
\end{align}

where $\mathcal{L}$ is the loss from Eq.~\eqref{eq:dice-loss} (see Sec.~\ref{subsec:surrogate-entropy}), $\hat{y}$ is the ground-truth segmentation mask, and $\mathbb{E}_{v \in y_{t+1}}[y_e(v)]$ denotes the expectation over the voxel-wise entropy values for all voxels included in $y_{t+1}$ from the region-growing operator. 
First two terms measure improvement in segmentation compared to ground truth between consecutive iterations and are collectively called the dice reward, while the final term encourages exploration in high-entropy (low-confidence) regions called the entropy reward, where $y_e$ is the entropy map obtained from the region-growing operator. This entropy reward encourages exploration of high-entropy areas i.e., those where the surrogate network is uncertain (voxel classification probabilities close to 0.5 leading to high expected values of $y_e(v)$). The hyper-parameter $\beta$ controls balance between exploitation and exploration, where a higher $\beta$ for exploration allows escaping local minima caused by dataset-level trends. 
Note that the first term rewards improved overlap (decrease in Dice loss), while the entropy bonus encourages exploration of uncertain regions.\\

\noindent\textbf{Optimisation} 
The agent network $h(\cdot; \phi)$ is trained to learn the optimal policy parameters $\phi^*$ that maximise the expected cumulative discounted reward over a finite time horizon $T$. The optimisation objective is defined as:
\begin{align}
\phi^* = \arg\max_{\phi} \, \mathbb{E}\left[\sum_{t=0}^{T-1} \gamma^t R_t \right],
\end{align}
where $\gamma \in [0,1)$ is the discount factor controlling the trade-off between immediate and future rewards. 
The expectation is taken over the state and action trajectories induced by the policy $h(\cdot; \phi)$ lasting up to time-step $T$.
In our implementation, the optimisation is performed using a policy-gradient based approach \cite{Schulman2017PPO}, allowing the network to iteratively improve its seed-point selection strategy to maximise the long-term segmentation performance. The training is summarised in Algo. \ref{algo:rl_train}.\\

\noindent\textbf{Inference to obtain the final segmentation} 
After training, the optimal policy $h(\cdot; \phi^*)$ is used to iteratively predict seed locations until convergence. 
After the initial user point prompt $v_{s,0}$, at each time step the agent selects the next seed voxel $v_{s,t}$, which is passed to the region-growing operator $g(\cdot)$ to update the segmentation.
The process terminates when the segmentation mask stabilises, defined as no further change in $y_t$ or upon reaching a maximum number of iterations $T$. The final segmentation is obtained as $y_T = g(x, v_{s,T})$, where $v_{s,T}$ is the final seed voxel selected by the trained agent and $y_T$ is the corresponding segmentation. 

At each decision step, the agent predicts a new seed voxel, from which region growing is re-initialised to generate a new segmentation mask. The newly generated mask replaces the previous segmentation rather than being accumulated, allowing the agent to correct prior over- or under-segmentation and preventing monotonic region growth. Although region growing uses fixed parameters, adaptability is achieved through iterative seed relocation guided by the RL policy and voxel-wise entropy, enabling sample-specific refinement.

\begin{algorithm}[!ht]
\caption{RL training for promptable segmentation}
\label{algo:rl_train}
\SetAlgoLined
\KwData{Image-label pairs $\{(x_i, \hat{y}_i)\}_{i=1}^N$, trained surrogate network $f(\cdot;\theta^*)$}
\KwResult{Trained agent $h(\cdot;\phi^*)$}
\BlankLine
\While{not converged}{
    Sample $(x,\hat{y})$ from the training set\;
    Obtain entropy map $y_e = f(x;\theta^*)$\;
    For time step $t = 0$, initialise heuristic seed point $v_{s,0}$ \;
    Apply region-growing for segmentation $y_0 = g(x, v_{s,0})$\;
    Set state $s_0 = (x, y_0)$\;

    \For{$t \leftarrow 0$ \KwTo $T-1$}{
        Select next action using agent $h(s_t;\phi)=  v_{s,t} = a_t $\;
        Apply region growing to get $y_{t+1} = g(x, v_{s,t})$\;
        Set next state $s_{t+1} = (x, y_{t+1})$\;
        Compute reward $R_t = \mathcal{L}(y_t, \hat{y})  - \mathcal{L}(y_{t+1}, \hat{y}) 
+ \beta\, \mathbb{E}_{v \in y_{t+1}}[y_e(v)]$\;
        Store transition $(s_t, a_t, R_t, s_{t+1})$ in buffer\;
        \If{$\|y_{t+1}-y_t\|_1 = 0$}{
            \textbf{break} 
        }
    }
    Update $\phi$ using $\arg\max_{\phi} \, \mathbb{E}\left[\sum_{t=0}^{T-1} \gamma^t R_t \right]$\;
}
\end{algorithm}


\section{Experiments}\label{sec:exp}

\subsection{Datasets}

\noindent\textbf{PROMIS:} 
\cite{Ahmed2017PROMIS} consists of multi-parametric MR images from 566 patients with suspected prostate cancer. T2-weighted (T2W), diffusion-weighted (DWI), and apparent diffusion coefficient (ADC) sequences form separate channels. Images were centre-cropped and resampled to $128 \times 128 \times 128$ voxels with intensities normalised. Voxel-level annotations for suspected cancerous regions, conducted by imaging researchers with agreed consensus, served as ground truth, with negative cases having fully-negative segmentation masks. During training initial seed points were sampled randomly within the lesion. For inference, seed points were also randomly chosen within the lesion, to simulate inter-observer variability. Data was randomly split into development and holdout sets with ratio 80:20, where heldout samples were used to report performance. 

\noindent\textbf{PICAI:}
\cite{Saha2024PICAI_LancetOncol} consists of multi-parametric MR images from 1090 patients from multiple international centres. These images had the same channels as PROMIS and were also resampled and normalised to the same ranges. 
Expert annotations for labelled cases served as ground truth, similar to PROMIS, with negative cases having fully-negative segmentation masks. Seed point locations were sampled in the same manner as the PROMIS dataset.
The data was randomly split into the development and holdout sets with ratio 80:20. The heldout samples were used to report performance.

\subsection{Network architectures and hyper-parameters}

Our open-source implementation, with details and hyper-parameter settings, is available at: \url{github.com/JQ-Sakura/prostate-rl-segmentation}. The training was conducted on two Nvidia Tesla V100 GPUs, with surrogate network training and RL agent training lasting approximately 24 and 96 hours.

\noindent\textbf{Surrogate network:} 
Followed a 3D UNet architecture \cite{Ronneberger2015UNet} with 4 downsampling and 4 upsampling blocks. Hyper-parameters are reported in the supplementary materials.

\noindent\textbf{Agent:}
The agent had two parts, the actor and the critic and used the proximal policy optimisation algorithm for training \cite{Schulman2017PPO}. Both the actor and critic networks adopt a shared 3D convolutional encoder followed by three fully connected layers. Each layer is followed by group normalization and LeakyReLU activation. The encoder receives four-channel volumetric input, comprising the three MR modalities (T2W, DWI, ADC) and segmentation mask. Hyper-parameters are reported in the supplementary materials.

\subsection{Comparisons}

We compare our method (RL-PromptSeg) with commonly used segmentation methods, including SAM \cite{Kirillov2023SAM}, SAM2 \cite{ravi2024sam2}, SAM3 \cite{carion2025sam3}, MedSAM \cite{Ma2024MedSAM}, MedSAM2 \cite{ma2025medsam2}, MedSAM3 \cite{liu2025medsam3}, Combiner \cite{Yan2024Combiner}, T2-predictor \cite{yi2025t2}, Swin-UNeTr \cite{Hatamizadeh2022SwinUNETR}, UniverSeg \cite{Butoi2023UniverSeg}, UNet \cite{Ronneberger2015UNet}, nnUNet \cite{Isensee2021nnUNet} and DinoV3 \cite{simeoni2025dinov3}. We report Dice scores (mean $\pm$ standard deviation) on holdout sets, with statistical significance assessed using paired t-tests. For reference, human expert performance from a second reader (3 years experience) was measured on all 114 heldout samples in the PROMIS dataset, to estimate a human benchmark for performance. Promptable models (SAM, MedSAM, and UniverSeg) were fine-tuned using the development set. For SAM and MedSAM, only the segmentation head (mask decoder) was fine-tuned, with the backbone encoder kept frozen, using the Adam optimiser with an initial learning rate of $1e^{-5}$ and batch size of 64, following the recommended protocols in the original publications. UniverSeg was fine-tuned end-to-end using the Adam optimiser with an initial learning rate of $1e^{-6}$ and batch size of 128. All other compared methods were trained end-to-end following their original training protocols, initialising with pre-trained weights where applicable. 

\subsection{Ablations}

Our ablation studies include investigating impact of inclusion of certain components. We compare our RL-PromptSeg method to ablated versions: 1) omitting the initial seed point prompts by using region-growing without an initial seed (no prompting); 2) omitting the entropy-based reward (no entropy reward); and 3) using a single T2 MR sequence instead of the multi-parametric MR image (no multi-parametric input). The impact of the parameter $\beta$, which controls the exploration-exploitation trade-off, is also studied. Other hyper-parameters (reported in the supplementary materials) were set using a grid search, and had limited impact in preliminary evaluations. 

\section{Results}\label{sec:res}

\subsection{Comparisons}

Comparisons are presented in Tab. \ref{tab:comparison}. Our proposed RL-PromptSeg method outperforms the previous state-of-the-art fully-automated method Swin-UNeTr by 9.9\% and 8.9\% percentage points for the PROMIS and PICAI datasets, respectively (p-values 0.010 and 0.004). It also outperforms the previous best promptable method UniverSeg by 21.9\% and 21.5\% for the two datasets, respectively (p-values 0.002 and 0.001). Compared to a human observer, statistical significance was not found with a p-value of 0.14. The prompt selection time was 131s per case, averaged across 10 patient cases, compared to a full annotation time of 1093s per case, allowing for a 10$\times$ reduction in annotation time, while maintaining comparable performance to a human observer and significantly exceeding fully-automated performance. Standard deviation with respect to the user-provided prompt point, with point being within the lesion, was 0.028. We also report impact of off-target prompts in the supplementary materials.

It is interesting to note that the surrogate network alone performs substantially worse than the proposed RL-PromptSeg framework, highlighting that the reported performance gains arise from the iterative RL-driven prompt refinement rather than from the surrogate model itself. Further analysis of the contribution of individual components is provided in the ablation studies.

\begin{table}[!ht]
\centering
\begin{tabular}{ccc}
\hline
\textbf{Model} & \makecell{\textbf{PROMIS}\\(Dice)} & \makecell{\textbf{PICAI}\\(Dice)} \\
\hline
SAM & 0.236 $\pm$ 0.107           & 0.294 $\pm$ 0.132 \\
SAM2 & 0.254 $\pm$ 0.124           & 0.301 $\pm$ 0.137 \\
SAM3 & 0.312 $\pm$ 0.113           & 0.298 $\pm$ 0.142 \\
MedSAM & 0.267 $\pm$ 0.138        & 0.342 $\pm$ 0.142\\
MedSAM2 & 0.291 $\pm$ 0.129       & 0.363 $\pm$ 0.144\\
MedSAM3 & 0.323 $\pm$ 0.148        & 0.376 $\pm$ 0.134\\
Combiner & 0.330 $\pm$ 0.180      & 0.469 $\pm$ 0.156\\
T2-predictor & 0.339 $\pm$ 0.192  & 0.394 $\pm$ 0.141\\
UniverSeg & 0.307 $\pm$ 0.216     & 0.351 $\pm$ 0.154\\ 
UNet & 0.327 $\pm$ 0.198          & 0.426 $\pm$ 0.153\\
UNet (surrogate) & 0.346 $\pm$ 0.174          & 0.453 $\pm$ 0.141\\
nnUNet & 0.414 $\pm$ 0.201        & 0.461 $\pm$ 0.137\\ %
Swin-UNeTr & 0.427 $\pm$ 0.185    & 0.477 $\pm$ 0.133\\ %
DinoV3 & 0.318 $\pm$ 0.163    & 0.438 $\pm$ 0.138\\
\hline
Human & 0.538 $\pm$ 0.094 & -\\
\hline
\textbf{RL-PromptSeg} & \textbf{0.526 $\pm$ 0.112} & \textbf{0.566$\pm$ 0.139} \\
\hline
\end{tabular}
\caption{Performance comparison.}
\label{tab:comparison}
\end{table}

\subsection{Ablations}

The ablation studies are presented in Tab. \ref{tab:ablation}. RL-PromptSeg outperformed ablated variants across both datasets with margins of $>$5\% percentage points (all p-values $<$0.010). The largest improvement was observed compared to the no entropy reward variant, where the entropy reward was omitted. Performance, for this variant reduces approximately to the level of other fully-automated methods.


The impact of varying the exploration-exploitation trade-off hyper-parameter $\beta$ is outlined in Tab. \ref{tab:ablation}. The performance for the optimal setting of $\beta$ was higher than the other values, with statistical significance (all p-values$<$0.020).

\begin{table}[!ht]
\subfloat[Impact of removing method components]{
\begin{tabular}{ccc}
\hline
\textbf{Model} & \makecell{\textbf{PROMIS}\\(Dice)} & \makecell{\textbf{PICAI}\\(Dice)} \\
\hline
{No prompting} & {0.456 $\pm$ 0.122} & {0.464$\pm$ 0.134} \\
{No entropy reward} & {0.425 $\pm$ 0.172} & {0.441 $\pm$ 0.141} \\
{No multi-parametric} & {0.461 $\pm$ 0.183} & {0.465$\pm$ 0.156} \\
\hline
\textbf{All components} & \textbf{0.526 $\pm$ 0.112} & \textbf{0.566$\pm$ 0.139} \\
\hline
\end{tabular}
}
\quad
\subfloat[Impact of $\beta$]{
\begin{tabular}{cc}
\hline
\textbf{$\beta$} & \makecell{\textbf{PROMIS}\\(Dice)} \\
\hline
0.0 & {0.425 $\pm$ 0.172} \\
0.2 & {0.447 $\pm$ 0.163} \\
0.4 & {0.473 $\pm$ 0.179} \\
0.6 & {0.498 $\pm$ 0.121} \\
\textbf{0.8} & \textbf{0.526 $\pm$ 0.112}  \\
1.0 & {0.472 $\pm$ 0.181} \\
\hline
\end{tabular}}
\caption{Ablation studies.}
\label{tab:ablation}
\end{table}

\subsection{Qualitative results}

Predicted segmentations from RL-PromptSeg are presented in Fig. \ref{fig:samples}, showing over-segmentation in some cases, with a detailed analysis in the supplementary materials. 

\begin{figure}[!ht]
    \centering
    \includegraphics[width=0.8\linewidth]{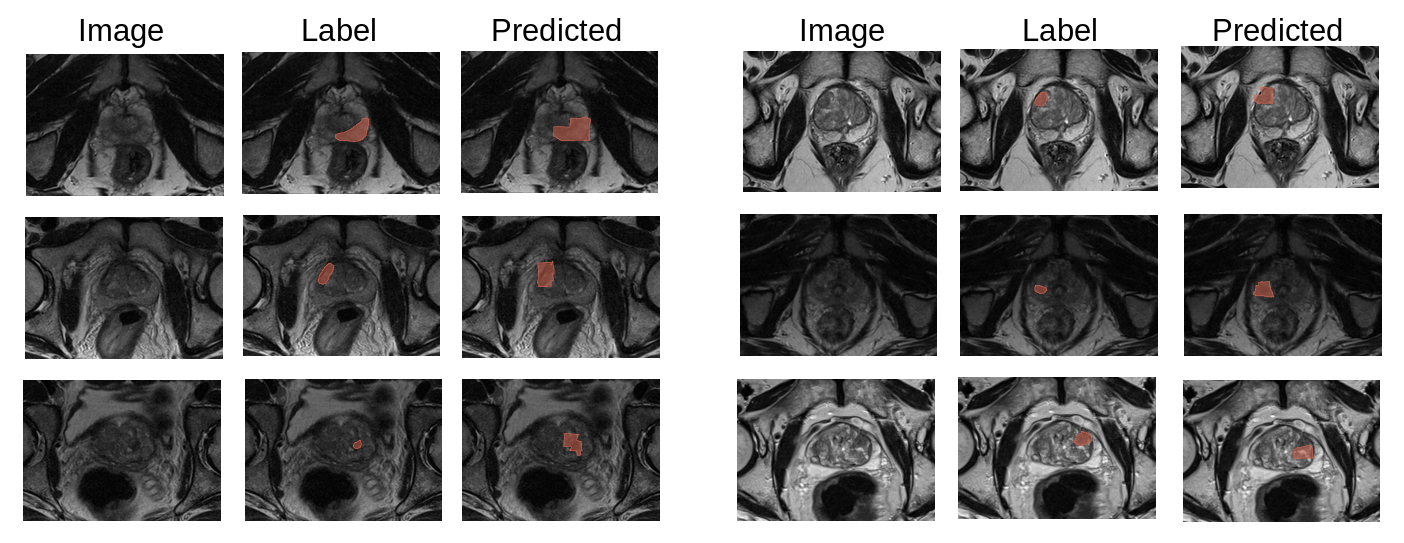}
    \caption{Samples from PROMIS segmented using our RL-PromptSeg approach.}
    \label{fig:samples}
\end{figure}

\section{Discussion}\label{sec:discussion}

These results demonstrate that our proposed framework enables effective prostate cancer segmentation on MR images. The method achieved performance comparable to expert radiologists and substantially exceeded existing fully-automated and promptable approaches across two independent datasets. A key advantage of our approach lies in its ability to combine the flexibility of prompting with sample-specific optimisation through RL. Unlike conventional deep learning models that rely on dataset-level trends, RL allows exploration for each sample, guided by the surrogate network’s voxel-wise entropy. The ablation experiments highlight the central role of the entropy-based reward in encouraging meaningful exploration, where its removal led to a marked decline in accuracy, reducing performance to that of conventional automated methods. Increasing the weight of the exploration-exploitation trade-off resulted in an improvement up to 0.8, where a further increase may necessitate much longer training times to accommodate added exploration. Our method significantly reduces the manual burden of prostate cancer segmentation, achieving a tenfold reduction in annotation time needing only a single point prompt per case, while maintaining radiologist-level accuracy. Despite its superior performance, we observed over-segmentation in some cases, however, this may be controlled by modifying region-growing hyper-parameters, or exploring alternative denser prompting strategies and point-to-segment algorithms. A full evaluation of inter-/ intra-operator variability and alignment with individual user intent would require prospective studies with multiple operators and interactive correction workflows, which we leave as important future work. Future work could also explore other problems where annotation variability causes poor performance. 

\section{Conclusion}\label{sec:conclusion}

We presented a reinforcement learning–based promptable segmentation framework for prostate cancer delineation on MR images. By modelling promptable segmentation as a sequential process, the method bridges the gap between manual and automated delineation by enabling sample-specific refinement guided by user prompts and voxel-wise entropy. The approach achieved radiologist-level performance while reducing annotation time tenfold, demonstrating its potential to accelerate clinical workflows, and dataset curation. Beyond MR, the formulation offers a general framework for different approaches to implement semi-automated segmentation across diverse tasks.

\section*{Acknowledgements / Funding}

This work is supported by the International Alliance for Cancer Early Detection, an alliance between Cancer Research UK [EDDAPA-2024/100014] \& [C73666/A31378], Canary Center at Stanford University, the University of Cambridge, OHSU Knight Cancer Institute, University College London and the University of Manchester.

\section*{Declarations}

\subsection*{Compliance with Ethical Standards}

\noindent\textbf{Competing interests:} All authors have no conflicts of interest to declare that are relevant to the content of this article.\\

\noindent\textbf{Ethics approval:} Our work uses open-source datasets where all original procedures performed in studies involving human participants were in accordance with the ethical standards of the institutional and/or national research committee and with the 1964 Helsinki Declaration and its later amendments or comparable ethical standards.\\

\noindent\textbf{Informed consent:} Informed consent was obtained from all individual participants included in the original studies.

\bibliography{sn-bibliography}

\end{document}


\title[Article Title]{Supplementary Material - Promptable segmentation with region exploration enables minimal-effort expert-level prostate cancer delineation}

\author[1]{\fnm{Junqing} \sur{Yang}}

\author[2]{\fnm{Natasha} \sur{Thorley}}

\author[3]{\fnm{Ahmed Nadeem} \sur{Abbasi}}

\author[2]{\fnm{Shonit} \sur{Punwani}}

\author[4]{\fnm{Zion} \sur{Tse}}

\author[1]{\fnm{Yipeng} \sur{Hu}}

\author*[1,4]{\fnm{Shaheer U.} \sur{Saeed}}\email{shaheer.saeed@qmul.ac.uk}

\affil[1]{\orgdiv{UCL Hawkes Institute; Department of Medical Physics and Biomedical Engineering}, \orgname{University College London}, \orgaddress{UK}}

\affil[2]{\orgdiv{Centre of Medical Imaging}, \orgname{University College London}, \orgaddress{UK}}

\affil[3]{\orgdiv{Department of Oncology}, \orgname{Aga Khan University Hospital}, \orgaddress{Pakistan}}

\affil[4]{\orgdiv{Centre for Bioengineering; School of Engineering and Materials Science}, \orgname{Queen Mary University of London}, \orgaddress{UK}}

\maketitle

\vspace{0.5cm}

\subsection*{Robustness analysis}

In clinical scenarios, with limited annotation time, point prompts may be placed near lesions but outside their exact boundaries. As such, to simulate this realistic scenario, we randomly sampled point prompt locations between 0-20 pixels or 0-10 mm away from lesion boundaries. For such perturbed points, the performance for RL-PromptSeg is compared with SAM \cite{Kirillov2023SAM}, SAM2 \cite{ravi2024sam2}, SAM3\cite{carion2025sam3} and MedSAM3 \cite{liu2025medsam3} (all other compared methods use no prompts, or bounding boxes, and thus are excluded from this experiment).

The impact of perturbed points for the PROMIS dataset is summarised in Tab. \ref{tab:robustness}. All methods showed poorer performance compared to points sampled within the lesion, however, RL-PromptSeg showed minimal performance reduction and had superior performance compared to other methods (p-values $<$0.001). Standard deviation with respect to perturbed point locations was 0.052. This robustness of RL-PromptSeg is attributable to the proposed RL formulation, which enables iterative exploration and prompt relocation rather than reliance on a single initial prompt location. The robustness of RL-PromptSeg to perturbed prompts provides a proxy for inter-/ intra-operator variability, demonstrating that the final segmentation is stable across different user point placements.

\begin{table}[!ht]
\centering
\begin{tabular}{ccc}
\hline
Model & \makecell{Prompts within lesion\\(Dice)} & \makecell{Prompts outside lesion \\(Dice)}\\
\hline
SAM & 0.236$\pm$0.107 & 0.162$\pm$0.097\\
SAM2 & 0.254$\pm$0.124 & 0.141$\pm$0.127\\
SAM3 & 0.312$\pm$0.113 & 0.234$\pm$0.115\\
MedSAM3 & 0.323$\pm$0.148 & 0.226$\pm$0.134\\
\hline
RL-PromptSeg & 0.526$\pm$0.112 & 0.473$\pm$0.146\\
\hline
\end{tabular}
\caption{Robustness analysis for PROMIS (perturbed point prompts outside lesion boundaries).}
\label{tab:robustness}
\end{table}

\subsection*{Generalisability analysis}

To assess the generalisability of the RL-PromptSeg framework to other applications we evaluate its performance on two other datasets: 1) LiTS: liver tumour segmentation \cite{bilic2023lits}; and 2) KiTS: kidney tumour segmentation \cite{heller2021kits}. The two datasets, consisting of 131 and 489 samples of CT images, are split into development and holdout sets with ratio 80:20, with performance reported on heldout samples. Only the best-performing general-purpose methods from the comparison on PROMIS and PICAI were included for comparison in this generalisability study, alongside the previous reported fully-automated best-performing methods CLIPSeg \cite{liu2023clip} and ASeg \cite{myronenko2023aseg} (methods that used LiTS or KiTS for pre-training were excluded).

The results for the generalisability study across the tasks of liver and kidney tumour segmentation on CT images are reported in Tab. \ref{tab:generalisability}. We observed statistically significant performance improvements for RL-PromptSeg compared to all other tested methods (all p-values $<$0.02). It is interesting to note that we also observed performance improvements compared to the best fully-automated methods for each application, albeit at the trade-off of increased inference-time compute for RL-PromptSeg in addition to the required initial user-provided prompt. These results indicate that the proposed RL-based promptable formulation generalises across organs and imaging modalities, suggesting applicability beyond prostate MR.

\begin{table}[!ht]
\centering
\begin{tabular}{ccc}
\hline
\textbf{Model} & \makecell{\textbf{LiTS}\\(Dice)} & \makecell{\textbf{KiTS}\\(Dice)} \\
\hline
SAM3        & 0.651 $\pm$ 0.101    & 0.613 $\pm$ 0.071 \\
MedSAM3     & 0.703 $\pm$ 0.099    & 0.724 $\pm$ 0.074\\
UNet        & 0.742 $\pm$ 0.097    & 0.718 $\pm$ 0.064\\
nnUNet      & 0.748 $\pm$ 0.085    & 0.720 $\pm$ 0.059\\ %
Swin-UNeTr  & 0.758 $\pm$ 0.093    & 0.714 $\pm$ 0.072\\ %
DinoV3      & 0.745 $\pm$ 0.103    & 0.727 $\pm$ 0.068\\
CLIPSeg     & 0.794 $\pm$ 0.081    & -\\
ASeg        & -                    & 0.764 $\pm$ 0.055\\
\hline
\textbf{RL-PromptSeg} & \textbf{0.803 $\pm$ 0.105} & \textbf{0.772$\pm$ 0.067} \\
\hline
\end{tabular}
\caption{Performance comparison for liver and kidney tumour segmentation.}
\label{tab:generalisability}
\end{table}

\FloatBarrier

\subsection*{Over-segmentation analysis}

Tab. \ref{tab:overseg} quantifies over-segmentation, using voxel-wise false positive rate (FPR), which directly measures excess background inclusion, alongside sensitivity to assess the trade-off between lesion coverage and over-segmentation. Only the best-performing methods from the above analysis were included in the comparisons. Note that specificity and false negative rate are not reported as they can be directly derived from the reported quantities. The results show that all automated methods exhibit higher over-segmentation than the human observer; however, RL-PromptSeg achieves substantially lower FPR than other automated approaches while maintaining sensitivity comparable to human performance.

\begin{table}[!ht]
\centering
\caption{Quantitative analysis of over-segmentation on the PROMIS dataset.}
\begin{tabular}{ccc}
\hline
\textbf{Model} & \makecell{\textbf{PROMIS}\\ \textbf{(FPR)}} & \makecell{\textbf{PROMIS}\\ \textbf{(Sensitivity)}} \\
\hline
SAM3            & 0.322 & 0.421  \\
MedSAM3         & 0.306 & 0.412  \\
Combiner        & 0.218 & 0.498 \\
T2-predictor    & 0.213 & 0.507 \\
UniverSeg       & 0.264 & 0.516  \\
Swin-UNeTr      & 0.281 & 0.523  \\
Human           & 0.133 & 0.571  \\
\hline
\textbf{RL-PromptSeg} & \textbf{0.144} & \textbf{0.569} \\
\hline
\end{tabular}
\label{tab:overseg}
\end{table}

\subsection*{Handling negative cases}

For negative cases, initial prompts are sampled within the prostate gland. A scan is classified as negative if the final segmentation does not exceed a minimum size threshold corresponding to the region-growing neighbourhood, suppressing small erroneous false-positive regions. Negative cases naturally yield minimal Dice improvement, discouraging region expansion through the reward formulation, without requiring an explicit ‘no-lesion’ action. In cases with multiple lesions, one prompt is used per lesion; such cases were infrequent in the evaluated datasets.

The patient-level sensitivity and specificity (with 95\% confidence intervals) for a human observer were reported as 0.88 (0.84–0.91) and 0.45 (0.39–0.51) in the PROMIS study \cite{Ahmed2017PROMIS}. This corresponds to classifying a single scan as positive or negative. For the same task on the holdout set in PROMIS, RL-PromptSeg achieved 0.84 accuracy with sensitivity and specificity of 0.84 (0.81–0.89) and 0.48 (0.40–0.53), which is in line with reported human performance.

\subsection*{Hyper-parameter settings}

The values of relevant hyper-parameters are summarised in Tab. \ref{tab:hyps}.

\begin{table}[!ht]
\centering
\begin{tabular}{c c}
\hline
Hyper-parameter & Value \\
\hline
Surrogate learning rate & $1e^{-4}$\\
Surrogate learning rate annealing & Cosine up-to $1e^{-6}$\\
Surrogate optimiser & Adam\\
Surrogate batch size & 256\\
Surrogate epochs & 200\\
\hline
$\beta$: reward scaling / exploration-exploitation & 0.8\\
$\tau_e$: entropy threshold & 0.1\\
$\tau_\sigma$: region growing intensity St.D. threshold & 0.3\\
\hline
RL learning rate & $1e^{-4}$\\
RL $\gamma$ & 0.99\\
RL batch size & 128\\
RL steps & 1000*512\\
RL clip range epsilon & 0.2\\
RL advantage lambda & 0.95\\
RL entropy coefficient & 0.01\\
\hline
\end{tabular}
\caption{Hyper-parameter values.}
\label{tab:hyps}
\end{table}

\bibliography{sn-bibliography}